\documentclass[12pt]{JHEP3}
\usepackage{amsmath,amsfonts,amsbsy}
\usepackage[dvips]{graphicx,epsfig}

\usepackage{mathrsfs}
\usepackage{amsmath,amssymb}
\usepackage{epsfig}
\usepackage{relsize}
\usepackage{graphicx}
\input epsf

\usepackage{epsfig}
\usepackage{relsize}

\input epsf

\usepackage{mathrsfs}
\usepackage{amsmath,amssymb}
\usepackage{epsfig}
\input epsf

\newcommand{\alg}[1]{\mathfrak{#1}}

 \csname
@addtoreset\endcsname{equation}{section}

\def\bec{\begin{center}}
\def\ec{\end{center}}
\def\a{\alpha}  

\def\b{\beta}

\def\s{\sigma}

\def\del{\partial}

\let\la=\label

\def\tr{{\rm tr}}

 \def\det{{\rm det\,}}
\def\be{\begin{equation}}
\def\ee{\end{equation}}
\def\bea{\begin{eqnarray}}
\def\eea{\end{eqnarray}}
\def\ba{\begin{array}}
\def\ea{\end{array}}

\def\CP{\mathbb{ CP}}

\def\osp{\alg{osp}(2,2|6)}

\newcommand{\ads}{{\rm  AdS}}

\def \B {\text B}

\def\a {\alpha}
\def\b {\beta}
\def\s {\sigma}
\def\pa {\partial}

\def\gb {g_{_{\rm B}}}
\def\go {g_{_{\rm O}}}
\def\gc _{g_{_{\chi}}}

\author{Luis F. Alday$^a$,\,
Gleb Arutyunov$^a${}\footnote{Correspondent fellow at Steklov
Mathematical Institute, Moscow.},\, Dmitri Bykov$^{b,c}$
 \\ $^{a}$ {\it Institute for Theoretical
Physics and Spinoza Institute,\\ ~~Utrecht University, 3508 TD
Utrecht, The Netherlands} \\ $^b$ {\it School of Mathematics,
Trinity College, Dublin 2, Ireland} \\ $^c$ {\it Steklov
Mathematical Institute, Moscow, Russia}\footnote{Emails:
L.F.Alday@uu.nl, G.Arutyunov@uu.nl, dbykov@maths.tcd.ie,
}}

\abstract{We derive the one-loop correction to the space-time
energy of a folded string in $\ads_4\times \CP^3$ carrying spin
$S$ in $\ads_4$ and angular momentum $J$ in $\CP^3$ in the long
string approximation. From this general result in the limit $J\ll
\log S $ we obtain the one-loop correction to the cusp anomalous
dimension which turns out to be  $-\frac{5\log 2}{2\pi}$. This
value appears to be in conflict with the prediction from the
recently conjectured all-loop Bethe ansatz.

 }

\title{Semiclassical Quantization of\\
Spinning Strings in {\bf AdS}$_4\times {\bf CP}^3$}

\preprint{
         \smaller{\smaller{\smaller{ITP-UU-08-51}}}\\[-.5ex]
          \smaller{\smaller{\smaller{SPIN-08-41}}}\\[-.5ex]
\          \smaller{\smaller{\smaller{TCDMATH 08-12}}}}

\begin{document}

\section{Introduction and summary}
Recently a new example of the AdS/CFT dual pair has been
conjectured \cite{Aharony:2008ug}. It involves the
three-dimensional ${\cal N}=6$ superconformal Chern-Simons theory
with  gauge group ${\rm SU}(N)\times {\rm SU}(N)$ and the theory
of M2-branes in the eleven-dimensional geometry $\ads_4\times {\rm
S}^7/{\mathbb Z}_k$, where $k$ is the Chern-Simons level. In the
scaling limit $k,N\to\infty$ with $\lambda=2\pi^2 N/k$ fixed, the
corresponding M-theory is effectively described by type IIA
strings moving in the $\ads_4 \times \CP^3$ background.

\smallskip

As shown in \cite{Arutyunov:2008if,Stefanski:2008ik}, the
Green-Schwarz action for type IIA string theory on \mbox{$\ads_4
\times \CP^3$} with $\kappa$-symmetry partially fixed can be
understood as the coset sigma-model on the same space supplied
with a proper Wess-Zumino term\footnote{See also \cite{Fre:2008qc}
for the sigma model description in the pure spinor formalism.}.
Indeed, type IIA supersting involves 32 fermionic degrees of
freedom (two Majorana-Weyl fermions in ten dimensions of opposite
chirality); due to $\kappa$-symmetry only 16 of them are physical.
On the other hand, the sigma model based on the coset space ${\rm
OSP }(2,2|6)/{\rm SO}(3,1)\times {\rm U}(3)$ contains 24 fermions.
However, it also exhibits $\kappa$-symmetry, which for generic
backgrounds allows one to gauge away precisely 8 fermions. The
remaining 16 fermions together with their bosonic partners render
the physical content of \mbox{$\ads_4 \times \CP^3$} superstring.
The coset sigma model is classically integrable
\cite{Arutyunov:2008if,Stefanski:2008ik} which opens a way to
investigate its dynamics in a way similar to the case of
$\ads_5\times {\rm S}^5$ superstring. In particular, an algebraic
curve encoding the solutions of the classical \mbox{$\ads_4 \times
\CP^3$} sigma model has been derived in \cite{Gromov:2008bz}.
Further aspects\footnote{We also point out \cite{Grignani:2008is},
where the Penrose limit, the Landau-Lifshitz limit, the dispersion
relation and giant magnon solutions both in infinite and finite
volume have been studied for the \mbox{$\ads_4 \times \CP^3$}
sigma model.} of classical integrability have been investigated in
\cite{Ahn:2008hj}.

\smallskip

Complementary, the planar superconformal Chern-Simons theory
appears to be integrable at leading order in the weak coupling
expansion \cite{Minahan:2008hf,Gaiotto:2008cg,Bak:2008cp}. The
corresponding Bethe equations can be embedded into those based on
the supergroup ${\rm OSP }(2,2|6)$ which provides a convenient
starting point to generalize them to higher loops. This, in
conjunction with the knowledge of the sigma model algebraic curve
and experience with the $\ads_5\times {\rm S}^5$ case
\cite{Kazakov:2004qf}, enabled the authors of \cite{Gromov:2008qe}
to conjecture the all-loop Bethe ansatz which should encode the
anomalous dimensions of gauge theory operators (string states) for
all values of $\lambda$. Some tests of the conjectured all-loop
Bethe equations have been carried out in
\cite{Astolfi:2008ji,Shenderovich:2008bs}. Finally, the
$\alg{su}(2|2)$-invariant S-matrix underlying these Bethe
equations has been identified \cite{Ahn:2008aa}.

\smallskip

In spite of these interesting developments, the question about
quantum integrability of the \mbox{$\ads_4 \times \CP^3$} sigma
model remains open. We would like to stress an apparent difference
with the $\ads_5\times {\rm S}^5$ model. In the latter case the
corresponding bosonic sigma model is quantum integrable and this
quantum integrability extends to the whole model including
fermions. In the present case, the bosonic model is integrable as
well, but quantum corrections to $\CP^3$ are known to spoil its
classical integrability \cite{Abdalla}. Thus, quantum
integrability of the full \mbox{$\ads_4 \times \CP^3$} model, if
exists, should essentially rely on inclusion of fermionic degrees
of freedom.

\smallskip

An important tool to investigate the question about quantum
integrability is provided by semiclassical quantization of rigid
string solutions \cite{Frolov:2002av,Frolov:2003qc}. Starting from
a simple classical string configuration, one finds the spectrum of
fluctuations around it. Summing up the fluctuation energies gives
the one-loop correction to the classical energy of the spinning
string which can be then compared to the value predicted by the
Bethe ansatz. A particularly convenient $\ads_5\times {\rm S}^5$
solution allowing for an explicit evaluation of the one-loop
energy correction is given by a rigid folded string carrying
Lorentz spin $S$ along $\ads_5$.  In the long string approximation
 the corresponding correction to the energy scales
logarithmically  with $S$ and is found \cite{Frolov:2002av} to be
$$
\delta E =-\frac{3\log 2}{\pi}\log S\,
.
$$
On the gauge theory side, this string solution corresponds to
twist two operators with large Lorentz spin $S$, for which the
difference between the scaling dimension $\Delta$ and spin $S$
behaves as
$$
\Delta-S=f(\lambda)\log S\, ,
$$
where the function $f(\lambda)$ is the universal scaling function,
also known as the cusp anomaly
\cite{Gubser:2002tv,Belitsky:2006en}. The quantity $\delta E$
provides the first correction to the strong coupling value of the
cusp anomaly which has been shown \cite{Benna:2006nd} to perfectly
agree with the Bethe ansatz prediction based on the BES equation
\cite{Beisert:2006ez}. The one-loop correction to the long
$(S,J)$-string, which in addition to the Lorentz spin $S$ also
carries angular momentum $J$ along a big circle of five-sphere,
has been obtained in \cite{Frolov:2006qe} and it provides the
leading strong coupling correction to the so-called generalized
scaling function $f(\lambda,J/\log S)$
\cite{Freyhult:2007pz,Fioravanti:2008rv}.

\smallskip

The purpose of the present paper is to perform a similar
semiclassical quantization of a rigid string spinning in the
$\ads_4 \times \CP^3$ space-time and obtain the corresponding
one-loop energy shift. Namely, we consider a rigid folded string
with Lorentz spin $S$ and angular momentum $J$ along a circle
${\rm S}^1\subset \CP^3$. The gauge theory operators dual to this
string solution are made of two bi-fundamental scalars with $S$
light-cone derivatives distributed among them, and they transform
in the irrep $[J,0,J]$ of $\alg{su}(4)$. By finding the
fluctuation spectrum around the classical solution in the long
string approximation, we obtain the corresponding one-loop energy
shift as a function of $S$ and $J$. In particular, in the limit of
"slow" rotation, $J\ll \log S$, we find that the corresponding
one-loop correction is given by
$$
\delta E =-\frac{5\log 2}{2\pi}\log S\, .
$$
Apparently, this result appears to be in contradiction with the
one conjectured in \cite{Gromov:2008bz}. According to
\cite{Gromov:2008bz}, the energy correction should be half of
that for the corresponding string solution on $\ads_5\times {\rm
S}^5$, {\it i.e.} it should be equal to $ -\frac{3\log
2}{2\pi}\log S$. The conjecture of \cite{Gromov:2008bz} was based
on the assumption that an unknown function $h(\lambda)$ entering
the all-loop Bethe ansatz has a vanishing subleading (constant)
term at strong coupling. Provided we adopt the same definition of
the cusp anomaly, we see that it is not the case. Clearly, further
investigations are needed to clarify this important issue.

\smallskip

We would like to stress that our computation  is a genuine
field-theoretic one and it does not rely on the knowledge of the
algebraic curve. It is done exclusively in the framework of the
coset sigma model. As observed in \cite{Arutyunov:2008if}, strings
which carry only AdS spin provide an example of a singular string
background, because the corresponding $\kappa$-symmetry
transformations instead of generic rank 8 have higher rank equal
to 12. Thus, to properly treat the fluctuation spectrum around
this singular solution, we keep throughout the calculations a
non-vanishing angular momentum $J$ which can be then regarded as
the regularization parameter. We find that the resulting
expression for the one-loop energy shift admits a smooth limit
$J\to 0$, which allows us to obtain the above stated result for
the cusp anomalous dimension of high spin operators.  Finally, we
notice that in the limit $J\ll \log S$ the fluctuation spectrum
contains 6 massless bosons and 2 massless fermions.  Thus, in
opposite to what happens in the $\ads_5\times {\rm S}^5$ case
\cite{Alday:2007mf} where only 5 bosonic massless excitations are
present, a would be "quantum bosonic $\CP^3$ model" is not
splitting off in this limit.

\smallskip

The paper is organized as follows. In the next section we discuss
the coset sigma model which captures the physics of type IIA
strings on \mbox{$\ads_4 \times \CP^3$}. In section 3 we present
the $(S,J)$-solution in terms of a coset element. Section 4 is
devoted to the analysis of the fluctuation spectrum around the
$(S,J)$-solution. Finally, in section 5 we compute the
corresponding one-loop energy shift. Appendix A contains the
details on the description of the coset space \mbox{$\ads_4 \times
\CP^3$}. In appendix B we provide a detailed treatment of
$\kappa$-symmetry transformations around the $(S,J)$-solution.

\smallskip

While preparing this manuscript for submission, the work
\cite{McLoughlin:2008ms} appeared, which seems to be in agreement
with our findings. We were also informed about
\cite{Krishnan:2008zs} where the same result for the one-loop
energy shift was obtained.

\section{String action}
The sigma model describing strings on the coset space
$\ads_4\times \CP^3
$
has been introduced in \cite{Arutyunov:2008if}. Denote by
$A=-g^{-1}dg$ the flat current constructed out of a coset
representative $g$.  The sigma model action reads as  \bea
\la{stac} S =-\frac{R^2}{4\pi \a'}\int {\rm d}\sigma{\rm d}\tau\,
{\mathscr L} \, ,\eea where $R$ is the radius of the AdS space and
the Lagrangian density is the sum of the kinetic and the
Wess-Zumino terms \bea \label{sLag} {\mathscr L}
=\gamma^{\a\b}{\rm str}\big(A^{(2)}_{\a}A^{(2)}_{\b}\big)+\kappa
\epsilon^{\a\beta}{\rm str}\big(A^{(1)}_{\a}A^{(3)}_{\beta}\big)\,
 .\eea
Here $A^{(k)}$ denotes a homogeneous component of $A$ of degree
$k$ under the ${\mathbb Z}_4$-automorphism $\Omega$ and
$\gamma^{\a\b}= h^{\a\b} \sqrt {-h}$ is the Weyl-invariant
combination of the world-sheet metric $h_{\a\beta}$ with
$\det\gamma=-1$. We also use the convention
$\epsilon^{\tau\sigma}=1$. To ensure $\kappa$-symmetry, the
parameter $\kappa$ in front of the Wess-Zumino term should be
equal to $\pm 1$.

\smallskip

To proceed, one has to chose an explicit parametrization of the
coset element $g$. We will pick up \bea\label{cosetrep} g=g_{_{\rm
O}}g_{_{\chi}}g_{_{\rm B}} ,\eea where $g=e^{\chi}$ depends on the
odd matrix $\chi$ comprising the 24 fermionic degrees of freedom
of the model. The element $g_{_{\rm O}}$ can be chosen in
different ways depending on which commuting isometries we would
like to be realized linearly. For instance, one can take\footnote{
For the definition of gamma-matrices, the $\alg{so}(6)$ Lie
algebra generators $T_{ij}$, the matrices $C_4,K_4,K,\Upsilon$ and
$T_i$ appearing throughout the paper see appendix A, and
\cite{Arutyunov:2008if}.} \bea\la{gads0} g_{_{\rm O}}
 &=&\left(\begin{array}{cc} e^{\frac{i}{2} t \Gamma^0 } & 0  \\
0~~ &~ e^{-\frac{\varphi}{2} (T_{34} + T_{56})}
\end{array}\right)\, , \eea
where $t$ and $\varphi$ are the global $\ads$ time and one of the
angles of $\CP^3$, respectively\footnote{In particular, the
algebra $\alg{so}(3,2)$ has rank two, so that one can choose the
diagonal matrices $\frac{i}{2}\Gamma^0$ and
$\frac{1}{4}[\Gamma^2,\Gamma^3]$ as the generators of commuting
isometries.}. Since the global symmetry group ${\rm OSP(2,2|4)}$
acts on $g$ from the left the isometries corresponding to
constants shifts of $t$ and $\phi$ will be realized linearly and
they do not act on the fermionic variables, {\it i.e.}, fermions
are unchanged under the corresponding ${\rm U}(1)$ transformations
\cite{Alday:2005jm}. We note that such a parametrization will be
suitable for imposition of the uniform light-cone gauge
\cite{Arutyunov:2004yx}.

\smallskip

Finally, the element $g_{_{\rm B}}$ comprises all the coordinates
parametrizing $\ads_4$ and $\CP^3$ except those which parametrize
the element $g_{_{\rm O}}$. Explicitly, \bea\la{gb} g_{_{\rm B}}
 &=&\left(\begin{array}{cc} g_{_{\ads}} & 0  \\
0~~ &~ g_{_{\CP}}
\end{array}\right)\, .\eea
The matrix $g_{_{\ads}}$ have the following characteristic
properties
$$
g_{_{\ads}}^tC_4g_{_{\ads}}C_4^{-1}={\mathbb I}\, ,
~~~K_4g_{_{\ads}}=g_{_{\ads}}^tK_4\,
,~~~~~\Gamma^5g_{_{\ads}}=g_{_{\ads}}^{-1}\Gamma^5\, ,
$$
Analogously, $ g_{_{\CP}}$ obeys the following requirements
$$
 g_{_{\CP}}^t g_{_{\CP}}={\mathbb I}\, , ~~~K_6  g_{_{\CP}}=
 g_{_{\CP}}^t K_6\, .
$$
As the consequence, the element $g_{_{\rm B}}$ satisfies the
following identity \bea\label{gbprop} \Upsilon g_{_{\rm B}}
\Upsilon^{-1}=g_{_{\rm B}}^{-1} \, ,\eea where $\Upsilon$ defines
an inner automorphism $\Omega$ of the complexified algebra $\osp$.
It is worth to point out that the matrix $g_{_{\rm O}}$ is
orthosymplectic  but it does not obey eq.(\ref{gbprop}) satisfied
by the element $g_{_{\rm B}}$.

\smallskip

As was explained in \cite{Alday:2005ww}, a convenient and compact
representation of the sigma model Lagrangian can be constructed in
terms of the following matrix $G$ \bea\label{G}
G=\left(\begin{array}{cc} g_{_{\ads}}K_4 g_{_{\ads}}^t &
0 \\
0 & g_{_{\CP}}K_6 g_{_{\CP}}^t
\end{array}\right)=\gb K \gb^t\, .
\eea By construction, this matrix is skew-symmetric: $G^t=-G$.
Introducing the split \bea\la{split} g_{_{\chi}}^{-1}g_{_{\rm
O}}^{-1}d(g_{_{\rm O}}g_{_{\chi}})={\rm F}+{\rm B}\, , \eea where
${\rm F}$ and ${\rm B}$ are odd and even superalgebra elements,
respectively, the Lagrangian (\ref{sLag}) can be cast in the form
\cite{Alday:2005ww,Arutyunov:2008if}
 \bea\nonumber  {\mathscr L}&=&\frac{1}{4}{\rm
str}\Big[\gamma^{\a\b} (\pa_{\a}GG^{-1}\pa_{\b}GG^{-1} + 4
\B_{\a}\pa_{\b}G G^{-1}+2\B_{\a}\B_{\b}+2
\B_{\a}G\B_{\b}^tG^{-1}) \\
~~~~~~~~~~~~~~~~~&&+ 2i\kappa\, \epsilon^{\a\b}{\rm F}_{\a} G{\rm
F}_{\b}^{st} G^{-1}\Big]\, . \la{Lor2}\eea The Lagrangian
(\ref{Lor2}) provides a convenient starting point for studying the
fluctuation spectrum around classical solutions of the string
sigma model.

\section{The $(S,J)$-string}
We choose as the background solution string spinning in the
directions $\phi$ and $\varphi$ of $\ads_4$ and $\CP^3$ spaces,
respectively. This naturally suggests to pick up as $g_{_{\rm O}}$
the following matrix \bea\la{gads1} g_{_{\rm O}}
 &=&\left(\begin{array}{cc} e^{\frac{i}{2}\, t\, \Gamma^0 -\frac{1}{4}\,\phi \, [\Gamma^2,\Gamma^3]} & 0  \\
0~~ &~ e^{-\frac{\varphi}{2} (T_{34} + T_{56})}
\end{array}\right)\, . \eea
Then, the AdS part of the element $g_{_{\rm B}}$ can be chosen as
follows \bea g_{_{\rm AdS}}= e^{\frac{i}{2}\rho\sin\psi \,
\Gamma^1-\frac{i}{2}\rho\cos\psi\, \Gamma^3}\, . \eea Hence, in
addition to the global time $t$, the space $\ads_4$ is
parametrized by the non-negative variable $\rho$ and by two
angles, $\phi$ and $\psi$. As to $g_{_{\rm CP}}$, since we
distinguish the angle $\phi$, it is convenient to choose the
remaining five coordinates on $\CP^3$ in the same way as was done
in \cite{Arutyunov:2008if}, namely, we parametrize $g_{_{\rm CP}}$
by one real coordinate $x_4$ and by two complex variables $y_{1}$
and $y_2$, see \cite{Arutyunov:2008if} for details. In order to
keep the present discussion clear, we refer the reader to appendix
A for the full details concerning the parametrization of $g_{_{\rm
CP}}$.

\smallskip

The background solution corresponding to the $(S,J)$-string can be
now obtained by putting to zero the AdS angle $\psi$ together with
the $\CP^3$ coordinates $x_4$ and $y_1,y_2$, and picking up the
rotating string ansatz for the remaining variables \bea
t=\varkappa \tau\, , ~~~~\phi=\omega_1\tau\,
,~~~~\varphi=\omega_2\tau\, , ~~~~\rho\equiv \rho(\sigma)\, . \eea
Of course, the spinning string ansatz  is embedded in the subspace
$\ads_3\times {\rm S}^1$ of $\ads_4\times \CP^3$, and, for this
reason, the corresponding solution must coincide with the one
obtained in \cite{Frolov:2002av}. We see that for the rotating
ansatz the components of $g_{_{\rm B}}$ reduce to \bea g_{_{\rm
AdS}}= e^{-\frac{i}{2}\rho\, \Gamma^3}\, ,
~~~~~~~g_{_{\CP}}=e^{\frac{\pi}{4}T_5}\,  \eea and, as the
consequence the coset element underlying the $(S,J)$-string
solution is of the form \bea g= \left(\begin{array}{cc}
e^{\frac{i}{2}\, \varkappa\tau \, \Gamma^0 -\frac{1}{2}\,\omega_1
\tau \,\Gamma^2\Gamma^3}
e^{-\frac{i}{2}\rho\, \Gamma^3} & 0  \\
0~~ &~ e^{-\frac{1}{2}\omega_2\tau (T_{34} +
T_{56})}e^{\frac{\pi}{4}T_5}
\end{array}\right)\, .
\eea In the next section we will use this representation to find
the Lagrangian for fluctuation modes.
\smallskip

Finally, we note that the parameters of the solution
$(\varkappa,\omega_1,\omega_2)$ are related to the Noether charges
of the model which are the space-time energy $E$, the $\ads$ spin
$S$, and the $\CP^3$ spin $J$ as follows \bea E&=&\sqrt{\lambda}\,
\varkappa \int \frac{{\rm d}\sigma}{2\pi}\, \cosh^2\rho \, ,
~~~~~S=\sqrt{\lambda}\, \omega_1\int \frac{{\rm d}\sigma}{2\pi}\,
\sinh^2\rho \, ,~~~J=\sqrt{\lambda}\, \omega_2 \, .\eea They are,
of course, the same as for the $(S,J)$-string spinning in
$\ads_3\times {\rm S^1 }$. Here the parameter $\lambda$ is related
to the AdS radius\footnote{Note that $\lambda$ is related to the
gauge theory parameters $k$ and $N$ as $\lambda=2\pi^2 N/k.$} as
$\sqrt{\lambda}=\frac{R^2}{\a'}$.

\smallskip

In this paper we are mostly interested in the so-called long
string limit corresponding to $\omega_1,\omega_2\to \infty$ with
the ratio $u=\frac{\omega_2}{\omega_1}= \frac{1}{\sqrt{1+x^2}}$
fixed. In this limit, \bea \kappa\approx \omega_1\, ~~~{\rm and
}~~~~x=\frac{\sqrt{\lambda}}{\pi J}\ln \frac{S}{J}~~~{\rm fixed}\,
.\eea The energy of the long string is then \bea
E=S+J\sqrt{1+x^2}+\ldots  \eea and it can be further approximated
by assuming the "fast" or "slow" limits which correspond to taking
$x\ll 1$ or $x\gg 1$, respectively \cite{Frolov:2006qe}.

\section{Lagrangian for quadratic fluctuations}

\subsection{Spectrum of bosonic fluctuations}
The Lagrangian for the quadratic fluctuations follows
straightforwardly from the bosonic part of the action
(\ref{Lor2}). In the conformal gauge we find
\begin{eqnarray}
\mathscr{L}^{(2)}_B=-\cosh^2\rho\,
\partial_\a \tilde{t} \partial^\a \tilde{t}+\sinh^2\rho\,  \partial_\a
\tilde{\phi}\partial^\a \tilde{\phi}+2\sinh{2\rho}\,
\tilde{\rho}(\varkappa
\partial_\tau \tilde{t}-\omega_1 \partial_\tau \tilde{\phi})+
\partial_\a \tilde{\rho} \partial^\a \tilde{\rho} + \nonumber \\ +
(\varkappa^2-\omega_1^2)\cosh{2\rho}\,
\tilde{\rho}^2+\sinh^2{\rho}\, (\partial_\a \psi \partial^\a
\psi+\omega_1^2 \psi^2)+\partial_\a \tilde{\varphi}
\partial^\a \tilde{\varphi}\nonumber\\+
\partial_\a x \partial^\a
x +\omega_2^2 x^2+ \partial_\a v_r
\partial^\a \bar{v}_r+\frac{\omega_2^2}{4} v_r \bar{v}_r
\, .\nonumber
\end{eqnarray}

We see that the part of the action for the $\ads_4$ fields
$\tilde{t},\tilde{\rho},\tilde{\phi},\psi$ and the angular $\CP^3$
variable $\tilde{\varphi}$, shown in the first two lines, exactly
agree with those in equation (5.10) of \cite{Frolov:2002av}. In
addition we have five $\CP^3$ fields, $x$ and two complex fields
$v_r$. Furthermore, the linearized Virasoro constraints read
\begin{eqnarray}
\frac{1}{2}(\omega_1^2-\varkappa^2)\sinh{2\rho}\tilde{\rho}-\varkappa
\cosh^2{\rho}\partial_\tau \tilde{t}+\omega_2
\partial_\tau \tilde{\varphi}+\omega_1 \sinh^2{\rho}\partial_\tau
\tilde{\phi}+\rho'~\partial_\sigma \tilde{\rho} \approx 0\, ,\\
\rho'~\partial_\tau \tilde{\rho}-\varkappa \cosh^2\rho
\partial_\sigma \tilde{t}+\omega_2 \partial_\sigma
\tilde{\varphi}+\omega_1 \sinh^2\rho \partial_\sigma \tilde{\phi}
\approx 0\, .
\end{eqnarray}

Obviously, the linearized Virasoro constraints are identical to
equations (5.11) and (5.12) of \cite{Frolov:2002av}. Then, the
physical fields from $\CP^3$ decouple completely from the rest. As
it can be seen from the Lagrangian and also as shown in
\cite{Arutyunov:2008if}, these are five massive fields. In units
of $\omega_2$, one of these fields have mass $m=1$ and the other
four $m=1/2$.

\smallskip

As for the other fields, in \cite{Frolov:2006qe} it was shown how
to compute the spectrum, in the long string limit, around the
solution we are interested it. According to \cite{Frolov:2006qe},
from the $\ads_4$ fields and $\tilde{\varphi}$ we get three
physical fields. One is $\psi$, with mass
$m_{\psi}^2=2\varkappa^2-\omega_2^2$, while the other two modes
have frequencies
\begin{eqnarray}
\Omega_{\pm n}^B=\sqrt{n^2+2\varkappa^2 \pm 2
\sqrt{\varkappa^4+n^2 \omega_2^2}},\hspace{0.3in}n=0,\pm1,\pm2,...
\end{eqnarray}
In the long string approximation $\varkappa \approx \omega_1$.
Notice that in the limit $\omega_1 \gg \omega_2$, we get one field
with mass (in units of $\omega_1$) $m^2=4$, one field with mass
$m^2=2$ and six massless fields, as opposed to the situation in
${\rm AdS}_5 \times {\rm S}^5$, where we get two fields of $m^2=2$
and five massless fields. It is these five massless fields which
give rise to the ${\rm O}(6)$ sigma model in this special limit
\cite{Alday:2007mf}. As we will see in the next section, in the
present case we will also find two massless fermions in this
limit. Hence the situation is pretty different to what happens in
${\rm AdS}_5 \times {\rm S}^5$.

\subsection{Spectrum of fermionic fluctuations}
\la{QFA} Here we will work out the spectrum of fermionic
fluctuations around the $(S,J)$-string. The relevant part of the
Lagrangian (\ref{Lor2}) contributing the quadratic action for
fermions is
\begin{equation}\label{0}
\mathscr{L}_{F}^{(2)}= {\rm str}\Big[\frac{1}{2}\gamma^{\alpha
\beta} {\rm B}_{\alpha}({\rm
B}_{\beta}+G\B_{\b}^tG^{-1})+\gamma^{\alpha \beta} {\rm
B}_{\alpha}
\partial_{\beta}G G^{-1}+{i\over 2} \kappa\,  \epsilon^{\alpha \beta} \, {\rm F}_{\alpha} \, G \, {\rm F}_{\beta}^{st}
G^{-1}\Big].
\end{equation}
According to the formula (\ref{split}), up to terms quadratic in
fermions, we have \bea {\rm F}&=& D\chi\, , ~~~~~ {\rm
B}=\go^{-1}d\go+\frac{1}{2} (D\chi \chi-\chi D\chi)\, , \eea where
we have introduced the covariant differential
$D\chi=d\chi+[\go^{-1}d\go,\chi]$. In the conformal
gauge\footnote{We take
$\gamma^{\tau\tau}=-1=-\gamma^{\sigma\sigma}$ and
$\gamma^{\tau\sigma}=0$.}  we, therefore, find the following
quadratic action  \bea \nonumber \mathscr{L}_{F}^{(2)}&=&
\frac{1}{2}{\rm
str}\Big[-(\go^{-1}\pa_{\tau}\go+G(\go^{-1}\pa_{\tau}\go)^tG^{-1})(D_{\tau}\chi\chi-\chi
D_{\tau}\chi)+\pa_{\sigma}G
G^{-1}(D_{\sigma}\chi\chi-\chi D_{\sigma}\chi)\Big]\\
&+& \frac{i}{2}\kappa\,  {\rm str}\Big[D_{\tau}\chi G
(D_{\sigma}\chi)^{st}G^{-1}-D_{\sigma}\chi G
(D_{\tau}\chi)^{st}G^{-1}\Big] \label{qf}\, ,\eea where we made
use of the fact that $\go$ and $G$ do not depend on $\sigma$ and
$\tau$, respectively. Explicitly, \bea \go^{-1}\pa_{\tau}\go=
\left(\begin{array}{cc} \frac{i}{2}\varkappa \Gamma^0
-\frac{1}{2}\omega_1\Gamma^2\Gamma^3\, &  0 \\
0  & -\frac{1}{2}\omega_2(T_{34}+T_{56})
\end{array}
\right) \eea and $\pa_{\sigma}GG^{-1}={\rm
diag}(-i\rho'\Gamma^3,0)$.

\medskip

It is clear that, in general, fermion masses will depend
non-trivially on the non-constant function $\rho(\sigma)$ and its
derivative  which enter in the above Lagrangian through the matrix
$G$. However, in the long string limit we are most interested in
here, one can approximate $\rho'(\sigma)\approx {\rm const}$.
Thus, in this limit one can attempt to redefine fermions as \bea
\chi \to W\chi W^{-1}\, , \la{fshift}\eea where the role of $W$
would be to remove the $\rho$-dependence from $G$. The matrix
$W\in {\rm OSP}(2,2|6)$ must satisfy a few natural requirements.
First, under redefinition (\ref{fshift}) the covariant
differential $D\chi$ undergoes the following transformation \bea
D\chi\to W \big(d\chi+[W^{-1} \go^{-1}d\go W,\chi]+[W^{-1}d
W,\chi]\big) W^{-1}\, . \label{cd}\eea Thus, if we do not want to
introduce an extra dependence on $\rho$, the matrix $W^{-1}d W$
should depend on the derivatives of $\rho'$ only and, when being
restricted to its AdS block, it should commute in the long string
limit with the corresponding block of $\go^{-1}\pa_{\tau}\go$. The
last requirement also guarantees that the kinetic term in
eq.(\ref{bL}) will not receive an extra $\rho$-dependence under
such redefinition of fermions. Second, $W$ must commute with
$\pa_{\sigma} G G^{-1}$, which is equivalent to the requirement of
commuting with $\Gamma^3$ (naturally embedded into $10\times
10$-matrices). This will ensure that the term in the Lagrangian
containing $\pa_{\sigma} G G^{-1}$ will not receive new
$\rho$-dependent terms. Finally, $W$ must be capable to remove
$\rho$ from  $G$, i.e the element $W^{-1} G (W^{t})^{-1}$ should
be independent of $\rho$. The conditions on $W$ stated above can
be satisfied {\it in the long string limit only} and they fix $W$
essentially uniquely. To construct $W$, we note that in the long
string limit $\varkappa\approx w_1$, so that the AdS part of
$\go^{-1}\pa_{\tau} \go$ becomes proportional to
$i\Gamma^0-\Gamma^2\Gamma^3\, $. Thus, we have to find an
$\alg{so}(3,2)$ Lie algebra element, such that it commutes with
two matrices
$$
i\Gamma^0-\Gamma^2\Gamma^3\, ~~~~~~~{\rm and}~~~~~~~\Gamma^3\, .
$$
One can easily see that the corresponding element is given by
$$
i\Gamma^3-\Gamma^0\Gamma^2\, .
$$
Here $\Gamma^3$ is the Lie algebra coset representative, while
$[\Gamma^0,\Gamma^2]$ belongs to the stability subalgebra
$\alg{so}(3,1)$. The last observation implies that taking $W$ in
the form \bea W=\left(\begin{array}{cc}
e^{-\frac{\rho}{2}(i\Gamma^3-\Gamma^0\Gamma^2)} &
0 \\
0 & e^{\frac{\pi}{4}T_5} \end{array}\right) \, ,\label{Wmat}\eea
we satisfy all the requirements stated, getting, in particular,
$$W^{-1} G (W^{-1})^{t}=K \, ,
$$
where the matrix $K$ is defined by eq.(\ref{mK}). Since
$e^{-\frac{\pi}{4}T_5}(T_{34}+T_{56})e^{\frac{\pi}{4}T_5}=-T_6$,
we see that after redefining the fermions by $W$, the covariant
derivative (\ref{cd}) acquires the following form
$$D_{\a}=\pa_{\a}+[Q_{\a},\ldots ],$$ where the composite vector field $Q_{\a}$
has the components \bea Q_{\tau}= {\rm diag} \Big(Q_{\tau}^{\rm
AdS},\, \, -\frac{1}{2}\omega_2 \, T_6\Big)\, ,
~~~~~~~~Q_{\s}=W^{-1}\pa_{\sigma}W  \, , \eea where \bea
 Q_{\tau}^{\rm AdS}&=& \frac{1}{4}\big( \varkappa
+\omega_1\big)\Big(i\Gamma^0 -\Gamma^2\Gamma^3\Big)+\\
&+&\frac{1}{4}(\varkappa-\omega_1)\Big[\cosh 2\rho\Big(i\Gamma^0 +
\Gamma^2\Gamma^3\Big) +\sinh 2\rho\Big(i\Gamma^2 +
\Gamma^0\Gamma^3\Big)\Big]\, . \nonumber \eea In the long string
limit $\varkappa\approx \omega_1$ the function $\rho$ drops out of
$Q_{\tau}$ as it should be. Also, by construction, in the long
string limit the commutator $[Q_{\a},Q_{\beta}]$ vanishes, {\it
i.e.} the connection $D_{\a}$ becomes flat.

\smallskip

We further note that since $\chi\in \osp$ its supertranspose is
$\chi^{st}=-C\chi C^{-1}$ and, therefore, after the redefinition
of fermions has been done, the action (\ref{qf}) can be cast in
the form \bea \nonumber \mathscr{L}_{F}^{(2)}&=& \frac{1}{2}\,
{\rm str}\Big[-(Q_{\tau}-\Upsilon
Q_{\tau}\Upsilon^{-1})(D_{\tau}\chi\chi-\chi
D_{\tau}\chi)+\pa_{\sigma}G
G^{-1}(D_{\sigma}\chi\chi-\chi D_{\sigma}\chi)\Big]+\\
&+& \frac{i}{2}\, \kappa\,  {\rm str}\Big[D_{\tau}\chi \Upsilon
D_{\sigma}\chi \Upsilon^{-1}-D_{\sigma}\chi \Upsilon D_{\tau}\chi
\Upsilon^{-1}\Big] \label{bL}\, .\eea Note that the kinetic term
for fermions is projected on the space ${\cal A}^{(2)}$ as
$Q_{\tau}-\Upsilon Q_{\tau}\Upsilon^{-1}\in {\mathcal A}^{(2)}$.
In particular, in the long string limit
$$
Q_{\tau}-\Upsilon Q_{\tau}\Upsilon^{-1}\approx i\varkappa
\Gamma^0+\omega_2 T_6\, .
$$
One can check that for a generic $\chi$ the kinetic term of the
Lagrangian  (\ref{bL}) is degenerate and it has rank 16. This is a
manifestation of $\kappa$-symmetry which allows one to remove 8
unphysical fermions out of 24 making thereby the kinetic term
non-degenerate \cite{Arutyunov:2008if}. As is shown in appendix B,
an admissible and convenient $\kappa$-symmetry gauge choice is
\bea \theta\, T_{56}=0 \, ,\eea which removes the fermions from
the fifth and the sixth column of $\chi$.

\smallskip

Introducing a 4 by 4 matrix $\vartheta$ made of non-vanishing
entries of $\theta$, we can write the quadratic $\kappa$-gauge
fixed Lagrangian in the long string approximation as follows
\bea\nonumber \mathscr{L}_{F}^{(2)}&=&-\varkappa\,  {\rm
tr}(\vartheta^t \Gamma^3\dot{\vartheta})
 +\frac{\varkappa^2}{2}{\rm tr}\Big[\vartheta^tC_4({\mathbb
 I}+i\Gamma^0\Gamma^2\Gamma^3)\vartheta\Big]
 +\frac{\omega_2^2}{4}{\rm tr}({\mathbb
I}-\Gamma^0)\vartheta^tC_4\vartheta\, \\
\la{MA} &-&\rho'{\rm
tr}(\vartheta^t\Gamma^0\vartheta')+\frac{\rho'^2}{2}\tr\Big[\vartheta^t
C_4
({\mathbb I}-i\Gamma^0\Gamma^2\Gamma^3)\vartheta \Big]\\
\nonumber &+&i\kappa \rho'\, \tr(\vartheta^t
\Gamma^0\Gamma^5\dot{\vartheta}K_4)+i\kappa\varkappa\,
\tr(\vartheta^t\Gamma^3\Gamma^5\vartheta'K_4)
+\kappa\varkappa\rho'\, \tr\Big[\vartheta^t({\mathbb
I}-i\Gamma^0\Gamma^2\Gamma^3)\Gamma^5\vartheta K_4\Big]\, .
 \eea
One can check that this action is hermitian provided the fermions
satisfy the reality condition
$\vartheta^\dagger=i\vartheta^t\Gamma^3$. Introducing the Dirac
conjugate
$\bar{\vartheta}=\vartheta^{\dagger}\Gamma^0=i\vartheta^t\Gamma^0\Gamma^3=-\vartheta^tC_4$,
we recognize that the reality condition is nothing else but the
Majorana condition.

\smallskip

To compute the one-loop energy shift, one has first to determine
the spectrum of fermion frequencies from the quadratic action
(\ref{MA}). This is essentially the same as to solve the
corresponding equations of motion. Every solution will be
characterized by the energy $k_0$ and the momentum $k_1$. Then,
every zero eigenvalue of the quadratic form defining (\ref{MA})
will give a (dispersion) relation between $k_0$ and $k_1$, while
the corresponding eigenstate will be a solution of the equations
of motion. Thus, we may look for the spectrum of the model by
requiring that the determinant of the corresponding quadratic form
is zero. There are as many particles in the theory as there are
linearly independent solutions.

\smallskip

This is precisely the strategy we would like to follow in this
paper, therefore let us discuss in more detail some subtleties
which we encounter on our way. Combining the fermions $\vartheta$
in one 16-dimensional vector, the action implied by (\ref{MA}) can
be schematically written as
\begin{equation}\label{action}
S=-\frac{R^2}{4\pi \a'}\int {\rm d}\sigma{\rm d}\tau\,
(\vartheta_{i}K^{ij}_{\tau}\del_{\tau}\vartheta_{j}+\vartheta_{i}K^{ij}_{\sigma}\del_{\sigma}
\vartheta_{j}+\vartheta_{i}M^{ij}\vartheta_{j})\, .
\end{equation}
In our treatment we will impose the reality condition on
$\vartheta$ only at the end of calculation, i.e. we prefer to
start with the action above, where in each term we have two
$\vartheta$'s rather than $\vartheta$ and $\vartheta^{\ast}$.
Varying the action, we get
\begin{eqnarray} \nonumber
\delta S&=&
 -\frac{R^2}{4\pi \a'}\int {\rm d}\sigma{\rm
d}\tau\,
\left(\delta\vartheta_{i}\widehat{K}^{ij}_{\tau}\del_{\tau}\vartheta_{j}+
\delta\vartheta_{i}\widehat{K}^{ij}_{\sigma}\del_{\sigma}\vartheta_{j}+
\delta\vartheta_{i}\widehat{M}^{ij}\vartheta_{j}\right),
\label{varaction}
\end{eqnarray}
where we have used anti-commutativity of fermions and integration
by parts. Here
\begin{eqnarray}
\nonumber
  \widehat{K}_{\tau} &=& K_{\tau}+K^{t}_{\tau}\, ,\\ \nonumber
  \widehat{K}_{\sigma} &=& K_{\sigma}+K^{t}_{\sigma}\, ,\\ \nonumber
  \widehat{M}&=& M-M^{t}.
\end{eqnarray}
Thus, equations for motion look as
\begin{equation}\label{eom}
(\widehat{K}_{\tau}\del_{\tau}+\widehat{K}_{\sigma}\del_{\sigma}+\widehat{M})\theta=0.
\end{equation}
In momentum space the equation above yields
\begin{equation}\label{eomom}
(i\widehat{K}
_{\tau}\,k_{0}+i\widehat{K}_{\sigma}\,k_{1}+\widehat{M})\theta=0
\, .
\end{equation}
As it follows from the discussion above, the spectrum of the model
is determined by the condition
\begin{equation}\label{spectr}
\mathscr{D}=\det\Big[i\widehat{K}_{\tau}\,k_{0}+i\widehat{K}_{\sigma}\,k_{1}+\widehat{M}\Big]=0,
\end{equation}
where $\widehat{K}_{\tau}$ and $\widehat{K}_{\sigma}$ are symmetric
matrices, and $\widehat{M}$ is antisymmetric. We view (\ref{spectr})
as an algebraic equation for $k_{0}$, and its roots (as functions of
$k_{1}$) give us the dispersion relations for all particles in the
theory.

\smallskip

Computing the determinant, we find \bea \nonumber
\mathscr{D}&=&2^8\omega_2^{16}\big[(2k_0-\omega_2)^2-4(k_1^2+\varkappa^2)\big]^2\big[(2k_0+\omega_2)^2-4(k_1^2+\varkappa^2)\big]^2
\times\\
&&~~~~~~~~~~~~~~~~~~~~~~~~~~~~\times
\big[k_0^4-k_0^2(2k_1^2+\varkappa^2)+k_1^2(k_1^2-\omega_2^2+\varkappa^2)\big]^2\,
. \eea Setting $k_{1}\equiv n \in \mathbb{Z}$, due to the fact
that this momentum corresponds to the compact $\sigma$ direction
of the string world-sheet, yields the following result for the
fermionic frequencies  (counting given in terms of elementary
fermions, rather than Majorana sets of fermions):
\begin{itemize}
  \item 2 fermions with frequency $\frac{\omega_2}{2}+\sqrt{n^{2}+\varkappa^{2}}\, $
  \item 2 fermions with frequency $-\frac{\omega_2}{2}+\sqrt{n^{2}+\varkappa^{2}}\, $
  \item 2 fermions with frequency $\sqrt{n^{2}+{1\over 2} \varkappa^{2}+{1 \over 2} \sqrt{\varkappa^{4}+4\omega_{2}^{2}n^{2}}}$
  \item 2 fermions with frequency $\sqrt{n^{2}+{1\over 2} \varkappa^{2}-{1 \over 2} \sqrt{\varkappa^{4}+4\omega_{2}^{2}n^{2}}}$
\end{itemize}
plus the other eight fermions whose frequencies are equal to the
above with negative sign. The reality condition then implies that
these negative frequency fermions are nothing else but the
conjugate momenta for the positive frequency ones. The constant
shifts by $\pm \omega_2/2$ in the first four frequencies can be
removed by an extra time-dependent redefinition of fermions,
similar to that done in \cite{Arutyunov:2008if}. The resulting
dispersion relation is the same as for relativistic fermions with
the mass $m^2=\varkappa^2$. In any case, even without doing this
field redefinition, the shifts by $\pm \omega_2/2$ are cancelled
out in the one-loop energy correction. In the special limit
$\varkappa\approx \omega_1\gg \omega_2$ the spectrum above will
contain two massless fermions.

\section{One-loop energy shift}
Having found the frequency modes of bosons and fermions, we can
readily compute the one-loop correction to the energy of the long
$(S,J)$-string. This  computation is very similar to that of
\cite{Frolov:2006qe}. The one-loop correction to the energy is
given by the following sum
\begin{eqnarray}
\nonumber \delta E =\frac{1}{\omega_1}\sum_{n=1}^\infty\left[\Big(
\Omega_{+,n}^B+\Omega_{-,n}^B+\sqrt{n^2+2\omega_1^2-\omega_2^2}+\sqrt{n^2+\omega_2^2}+4\sqrt{n^2+\frac{\omega_2^2}{4}}\,
\Big)-
\right.\\
-\left. \Big(2\Omega^F_{+n}+2\Omega^F_{-n}+4\sqrt{n^2+\omega_1^2}
\Big) \right]\, , \la{1loopE}
\end{eqnarray}
where
\begin{equation}
\Omega^F_{\pm,n}=\sqrt{n^2+\frac{\omega_1^2}{2}\pm
\frac{1}{2}\sqrt{\omega_1^4+4\omega_2^2 n^2}}\, ,
\end{equation}
are the non-trivial fermionic frequencies found in the previous
section. It is gratifying to see that the divergencies of bosons
cancel against those of fermions, so that the sum (\ref{1loopE})
is convergent.

\smallskip

We are most interested in the value of the sum in the scaling
limit, $\omega_1,\omega_2 \rightarrow \infty$ with
$u=\frac{\omega_2}{\omega_1}$ fixed. Following
\cite{Frolov:2006qe}, in this limit, the sum can be replaced by an
integral
\begin{eqnarray}
\nonumber \delta E=\omega_1 \int_0^\infty {\rm d}p \left[ \Big(
\Omega^B_+(p)+\Omega^B_-(p)+\sqrt{p^2+2-u^2}+\sqrt{p^2+u^2}+4\sqrt{p^2+\frac{u^2}{4}}
\Big)\right.-\\
- \left.\Big( 2\Omega^F_+(p)+2\Omega^F_-(p)+4 \sqrt{p^2+1}\Big)
\right]\, ,
\end{eqnarray}
where
\begin{eqnarray}
\Omega^B_{\pm}(p)=\sqrt{p^2+2 \pm 2 \sqrt{1+p^2
u^2}},\hspace{0.3in}\Omega^F_{\pm}(p)=\sqrt{p^2+\frac{1}{2} \pm
\frac{1}{2}\sqrt{1+4 p^2 u^2}}\, .
\end{eqnarray}
The simplest way to compute the integral is to impose a cut-off
and send it to infinity at the end of the computation. The
integrals involving $\Omega^B$ and $\Omega^F$ can be simplified by
using the identities
\begin{eqnarray}
\nonumber
\Omega^B_+(p)+\Omega^B_-(p)&=&\sqrt{4u^2+(p+\sqrt{p^2+4-4u^2})^2}\, \, ,\\
\nonumber
\Omega^F_+(p)+\Omega^F_-(p)&=&\frac{1}{2}\sqrt{4u^2+(2p+\sqrt{4p^2+4-4u^2})^2}\,
\,  .
\end{eqnarray}
Notice that these two expressions are related as
$$\Omega^B_+(p)+\Omega^B_-(p)=2\Omega^F_+(p/2)+2\Omega^F_-(p/2).$$
The integrals can then be easily performed by making the change of
variables $p \rightarrow q=p+\sqrt{p^2+4-4u^2}$. Finally, we
obtain
\begin{eqnarray}
\delta E= \frac{\omega_1}{4}
\Big[-2u^2\log{u^2}-2\log{(8-4u^2)}+u^2\log{(16(2-u^2))}+ \nonumber\\
+2\left(-1+u^2+\sqrt{1-u^2}+(u^2-2)\log{(1+\sqrt{1-u^2})}
\right)\Big]\, .
\end{eqnarray}
This formula describing the one-loop correction to the classical
energy of the long $(S,J)$-string is our main result. It has to be
compared with a corresponding result for the \mbox{$(S,J)$-string}
spinning in $\ads_5 \times {\rm S}^5$ given by eq.(2.29) in
\cite{Frolov:2006qe}. Curiously enough, we find that
\begin{equation}
\label{disagree} \delta E^{\ads_4 \times \CP^3}-\frac{1}{2} \delta
E^{\ads_5 \times {\rm S}^5}=\omega_1(u^2-1)\log{2}\, .
\end{equation}
This result is with an apparent contradiction to the conjecture
made in \cite{Gromov:2008qe}. According to their claim, the r.h.s.
of eq.(\ref{disagree}) should vanish. In the $u \to 0$ limit we
obtain
\begin{equation}
\delta E= -\frac{5}{2}\omega_2 x \log{2}=-\frac{5\log{2}}{2 \pi}\,
\log{\frac{S}{J}}\, .
\end{equation}
The coefficient in front of $\log{\frac{S}{J}}$ should be
interpreted as the one loop correction to the cusp anomalous
dimension.

\section*{Acknowledgements}
We thank Sergey Frolov and Arkady Tseytlin for many valuable
discussions. The work by L.~F.~A. was supported by VENI grant
680-47-113. The work of G.~A. was supported in part by the RFBI
grant N05-01-00758, by the grant NSh-672.2006.1, by NWO grant
047017015 and by the INTAS contract 03-51-6346. The work of D.B.
was supported in part by grant of RFBR ¹ 08-01-00281-a and in part
by grant for the Support of Leading Scientific Schools of Russia
NSh-795.2008.1. The work of G.~A. and D.~B. was supported in part
by the EU-RTN network {\it Constituents, Fundamental Forces and
Symmetries of the Universe} (MRTN-CT-2004-512194).

\appendix

\section{The coset space $\ads_4\times \CP^3$}
To make the paper self-contained, in this appendix we recapitulate
the basic facts about the description of the coset space
$\ads_4\times \CP^3={\rm OSP}(2,2|6)/{\rm SO}(3,1) \times {\rm
U}(3)$.

As in \cite{Arutyunov:2008if}, we introduce the following
gamma-matrices \bea \label{Gmatrices}
\begin{aligned} \Gamma^0&=&{\scriptsize \left(\begin{array}{cccc}
1 & 0 & 0 & 0 \\
0 & 1 & 0 & 0 \\
0 & 0 & -1 & 0 \\
0 & 0 & 0 & -1 \\
\end{array}\right)
}\, , ~~~~~ \Gamma^1={\scriptsize \left(\begin{array}{cccc}
0 & 0 & 1 & 0 \\
0 & 0 & 0 & -1 \\
-1 & 0 & 0 & 0 \\
0 & 1 & 0 & 0 \\
\end{array}\right)
}\, , \\
\Gamma^2&=&{\scriptsize \left(\begin{array}{cccc}
0 & 0 & 0 & 1 \\
0 & 0 & 1 & 0 \\
0 & -1 & 0 & 0 \\
-1 & 0 & 0 & 0 \\
\end{array}\right)
}\, , ~~~~~ \Gamma^3={\scriptsize \left(\begin{array}{cccc}
0 & 0 & 0 & -i \\
0 & 0 & i & 0 \\
0 & i & 0 & 0 \\
-i & 0 & 0 & 0 \\
\end{array}\right)
}\,  \end{aligned}
 \eea
satisfying the Clifford algebra relations
$\{\Gamma^{\mu},\Gamma^{\nu}\}=2\eta^{\mu\nu}$, where
$\eta^{\mu\nu}$ is Minkowski metric with signature $(1,-1,-1,-1)$.
We also define \bea
 \Gamma^5=-i\Gamma^0\Gamma^1\Gamma^2\Gamma^3\, ,
 ~~~~~C_4=i\Gamma^0\Gamma^3\, ,~~~~~~K_4=-\Gamma^1\Gamma^2\, .
\eea  In particular, $C_4$ is the charge conjugation matrix:
$(\Gamma^{\mu})^t=-C_4\Gamma^{\mu}C_4$, while $K_4$ satisfies
$(\Gamma^{\mu})^t=K_4\Gamma^{\mu}K_4^{-1}$. One has
$$
(\Gamma^5)^2={\mathbb I}\, , ~~~~~~K_4^2=-{\mathbb I}\,
,~~~~~C_4^2=-{\mathbb I}\, , ~~~~~~\Gamma^5=K_4C_4\, .
$$
The generators $\frac{1}{4}[\Gamma^{\mu},\Gamma^{\nu}]$ span the
algebra $\alg{so}(3,1)\sim \alg{usp}(2,2)$. Adding to this set of
generators the four matrices $\frac{i}{2}\Gamma^{\mu}$, one
obtains a realization of $\alg{so}(3,2)$.

\smallskip

Consider $10\times 10$ supermatrices
 \bea
A=\left(\begin{array}{ll} X ~&~ \theta \\
\eta ~&~ Y   \end{array}\right) \, ,\eea where $X$ and $Y$ are
even (bosonic) $4\times 4$ and $6\times 6$ matrices, respectively.
The $4\times 6$ matrix $\theta$ and the $6\times 4$ matrix $\eta$
are odd, i.e. linear in fermionic variables. As the matrix
superalgebra, the Lie superalgebra $\osp$, is spanned by
supermatrices $A$ satisfying two conditions \bea
A^{st}=-{C}A{C}^{-1}\, ,~~~~~A^{\dagger}=-\Gamma A{\Gamma}^{-1}\,
, \eea where $C={\rm diag}(C_4,{\mathbb I}_{6})$ and $\Gamma={\rm
diag}(\Gamma^0,-{\mathbb I}_{6})$. Here $A^{st}$ stands for the
supertranspose of $A$: \bea
A^{st}=\left(\begin{array}{rr} X^t ~&~-\eta^t \\
 \theta^t ~&~ Y^t   \end{array}\right) \, .\eea
The bosonic subalgebra of $\osp$ is
$\alg{usp}(2,2)\oplus\alg{so}(6)$. Explicitly, the fermionic
matrices obey \bea \eta=-\theta^t C_4\, ,
~~~~~~~~~\theta^*=i\Gamma^3 \theta\, ,\label{rel3}\eea i.e.
fermions are symplectic Majorana with the total number of real
fermionic components equal to 24.

\smallskip

We further introduce a $6\times 6$ matrix $K_6$ and a $10\times
10$ matrix $K={\rm diag}(K_4,K_6)$: \bea\la{mK} K_6={\mathbb
I}_3\otimes \left(\begin{array}{cc} 0 & 1 \\ -1 & 0
\end{array}\right)\, , \, ~~~~~~~
K={\mathbb I}_5\otimes \left(\begin{array}{cc} 0 & 1 \\ -1 & 0
\end{array}\right)\,
 . \eea This matrices can be used to define an
automorphism $\Omega$ of order four of the complexified algebra
$\alg{osp}(4|6)$ \bea
\Omega(A)=\left(\begin{array}{rr} K_4X^tK_4 ~&~ K_4\eta^tK_6 \\
-K_6\theta^tK_4 ~&~ K_6Y^tK_6\end{array}\right)=-\Sigma
KA^{st}K^{-1}\Sigma^{-1}\, . \eea Here $\Sigma={\rm diag}({\mathbb
I}_4,-{\mathbb I}_6)$ is the grading matrix. The orthosymplectic
condition for $A$ implies \bea \Omega(A)=(\Sigma K C) A (\Sigma K
C)^{-1}\equiv \Upsilon A \Upsilon^{-1}\, , \eea {\it i.e.}
$\Omega$ is an inner automorphism. Explicitly, \bea\la{Upsilon}
\Upsilon=\left(\begin{array}{cc} \Gamma^5 & 0 \\
0 & -K_6
\end{array}\right)\, . \eea

As the vector space, ${\cal A}=\osp$ can be decomposed under
$\Omega$ into the direct sum of homogeneous components: ${\cal
A}=\sum_{k=0}^3 {\cal A}^{(k)}$, where the projection $A^{(k)}$ of
a generic element $A\in \osp$ on the subspace ${\cal A}^{(k)}$ is
define as \bea
A^{(k)}=\frac{1}{4}\Big(A+i^{3k}\Omega(A)+i^{2k}\Omega^2(A)+i^k\Omega^3(A)\Big)\,
. \label{Z4proj} \eea In particular, ${\cal
A}^{(0)}=\alg{so}(3,1)\oplus {\rm u}(3)$.

\smallskip

Throughout  the paper we use the generators $T_{ij}$ of
$\alg{so}(6)$ defined as $T_{ij}=E_{ij}-E_{ji}$, where $E_{ij}$
are the standard matrix unities. We also introduce the following
six matrices $T_6$ which are Lie algebra generators of
$\alg{so}(6)$ along the $\CP^3$ directions: \bea
\label{Tmatrices}\begin{aligned}
T_1&=E_{13}-E_{31}-E_{24}+E_{42}\, ,
~~~~~~~~T_2=E_{14}-E_{41}+E_{23}-E_{32}\, , \\
T_3&=E_{15}-E_{51}-E_{26}+E_{62}\, ,
~~~~~~~~T_4=E_{16}-E_{61}+E_{25}-E_{52}\, , \\
T_5&=E_{35}-E_{53}-E_{46}+E_{64}\, ,
~~~~~~~~T_6=E_{36}-E_{63}+E_{45}-E_{54}\, .
\end{aligned}
 \eea
These generators are normalized as  ${\rm
tr}(T_iT_j)=-4\delta_{ij}$.

\smallskip

According to \cite{Arutyunov:2008if}, a generic ${\rm SO}(6)$
element parametrizing the coset space $\CP^3={\rm SO}(6)/{\rm
U}(3)$ can be written as \bea g_{_{\CP}}=e^{y_i T_i}\, . \eea

We parametrize  $\CP^3$ by means of  the spherical coordinates
$(r,\varphi,\theta,\a_1,\a_2,\a_3)$, or, alternatively, by means
of three complex inhomogeneous coordinate $w_i$, \bea\la{param}
y_1 + i y_2 &=& r\, \sin \theta \cos\frac{\alpha_1}{2}
\,e^{\frac{i}{2} (\alpha_2+\alpha_3)+\frac{i}{2}\varphi
}=\frac{r}{|w|}w_1 \,,\\\nonumber
   y_3 + i y_4 &=& r\, \sin \theta \sin
 \frac{\alpha_1}{2} \,e^{-\frac{i}{2}  (\alpha_2-\alpha_3)+\frac{i}{2} \varphi }=\frac{r}{|w|}w_2 \,,\\\nonumber
   y_5 + i y_6 &=& r\,  \cos \theta \, e^{i \varphi }=\frac{r}{|w|}w_3 \,
   ,
\eea where $|w|^2=\bar{w}_k w_k$ and $\sin
r=\frac{|w|}{\sqrt{1+|w|^2}}$. The geodesic circle described by
the angle $\varphi$ corresponds to taking $\theta=0$ and
$r=\frac{\pi}{4}$, or, equivalently, $w_3=e^{i\varphi}$ and
$w_1=0=w_2$. If we further extract a geodesic angle $\varphi$ by
introduce one real field $x$ and two complex fields $v_1$ and
$v_2$: \bea w_3=(1-x)e^{i\varphi}\, , ~~~w_1=\frac{1}{\sqrt{2}} \,
v_1e^{i\varphi/2}\, ,~~~~~~~w_2=\frac{1}{\sqrt{2}}\,
v_2e^{i\varphi/2}\, , \eea then the corresponding quadratic action
for the $\CP^3$ fluctuation modes around the $(S,J)$-string
solution coincides with the plane-wave action obtained in
\cite{Arutyunov:2008if}.

\section{Kappa-symmetry}
\smallskip

Here we present an independent analysis of $\kappa$-symmetry
transformations in the background of the $(S,J)$-string. As was
explained in \cite{Arutyunov:2008if}, $\kappa$-symmetry acts on
the coset element by multiplication from the right: \bea g\to
ge^{\epsilon}=g'g_c\, , \eea where $g_c$ is a compensating group
element from the denominator of the coset. We see that at linear
order in $\chi$ and $\epsilon$ we get \bea g\to g_{_{\rm
O}}g_{_{\chi}}g_{_{\rm B}}e^{\epsilon} = g_{_{\rm O}}e^{\chi}
e^{g_{_{\rm B}}\epsilon g_{_{\rm B}}^{-1}}\gb \approx g_{_{\rm O}}
e^{\chi+ g_{_{\rm B}}\epsilon g_{_{\rm B}}^{-1}} g_{_{\rm B}}\eea
Thus, at the linearized level the fermion matrix $\chi$ changes
under the $\kappa$-symmetry variation as  \bea \chi\to \chi+
g_{_{\rm B}}\epsilon g_{_{\rm B}}^{-1}\, . \eea Note also that the
compensation matrix $g_c$ which depends on the even number of
fermions does not arise for the linearized transformations. The
parameter $\epsilon=\epsilon^{(1)}+\epsilon^{(2)}$ in the above
formula is the one found in \cite{Arutyunov:2008if}, e.g., \bea
\hspace{-0.7cm}\epsilon^{(1)}=A_{\a,-}^{(2)}A_{\b,-}^{(2)}\kappa_{++}^{\a\b}
+\kappa_{++}^{\a\b}A_{\a,-}^{(2)}A_{\b,-}^{(2)}+A_{\a,-}^{(2)}\kappa_{++}^{\a\b}A_{\b,-}^{(2)}-\frac{1}{8}\,
{\rm str}(\Upsilon^2
A_{\a,-}^{(2)}A_{\b,-}^{(2)})\kappa_{++}^{\a\b}\, , \label{kappa}
\eea where $\kappa_{++}^{\a\b}$ is the $\kappa$-symmetry
parameter\footnote{We present the complete analysis for
$\epsilon^{(1)}$ only, the computation of $\epsilon^{(2)}$ goes
along the same lines.}. It is easy to find \bea
A^{(2)}_{\tau}&=&-\frac{1}{2}\gb^{-1}\Big(\go^{-1}\pa_{\tau}\go+G(\go^{-1}\pa_{\tau}\go)^t
G^{-1}\Big)\gb \, ,\\
A^{(2)}_{\sigma}&=&-\frac{1}{2}\gb^{-1}\big(\pa_{\sigma}G
G^{-1}\big)\gb \, .
 \eea
Hence, in the conformal gauge \bea
A^{(2)}_{\tau,-}=\frac{1}{2}(A^{(2)}_{\tau}-A^{(2)}_{\sigma})\equiv
\gb^{-1}\hat{A}\gb\, , \eea where \bea
\hat{A}=-\frac{1}{4}\Big(\go^{-1}\pa_{\tau}\go+G(\go^{-1}\pa_{\tau}\go)^t
G^{-1}-\pa_{\sigma}G G^{-1}\Big) \, .\eea An element $G$ entering
the last formula is determined from eq.(\ref{G}) to be
\bea\label{Gred} G=\left(\begin{array}{cc} e^{-\frac{i}{2}\rho \,
\Gamma^3}K_4 e^{-\frac{i}{2}\rho\, \Gamma^3} &
0 \\
0 & e^{\frac{\pi}{4}\, T_{5}}K_6 e^{-\frac{\pi}{4}\, T_{5}}
\end{array}\right)=\left(\begin{array}{cc} e^{-i\rho \, \Gamma^3}K_4
 &
0 \\
0 & e^{\frac{\pi}{2}\, T_{5}}K_6
\end{array}\right)\, .
\eea We also note that since we pulled out the factor $\gb$ out of
$A^{(2)}$, the matrix $\hat{A}$ is not element of the space ${\cal
A}^{(2)}$.

\smallskip

Thus, under $\kappa$-symmetry transformation the fermionic matrix
$\chi$ is shifted by
$$
g_{_{\rm B}}\epsilon^{(1)} g_{_{\rm B}}^{-1}=
\hat{A}\hat{A}\kappa+\kappa
\hat{A}\hat{A}+\hat{A}\kappa\hat{A}-\frac{1}{8}{\rm
str}(\Upsilon^2\hat{A}\hat{A})\kappa\, .
$$
In order find out implementation of this formula for $\chi$, we
have to understand the structure of the matrix $\hat{A}$.
Calculations reveal the following remarkably simple formula \bea
\hat{A}=\frac{\omega_2}{4} \left(\begin{array}{cc}
-\frac{i}{\omega_2}e^{-\frac{i}{2}\rho\Gamma^3}\big(\varkappa\cosh\rho\Gamma^0-\omega_1\sinh\rho
\Gamma^2+\rho'\Gamma^3\big)e^{\frac{i}{2}\rho\Gamma^3} &
0 \\
0 & T_{34}+T_{56}
\end{array}\right)\, .
\nonumber \eea The non-trivial Virasoro constraint written in
terms of $\hat{A}$ implies \bea\label{Virasoro} 0=4\, {\rm
str}(\hat{A}\hat{A})=\rho'^2+\omega_1^2\sinh^2\rho-\varkappa^2\cosh^2\rho+\omega_2^2\,
, \eea which is an equation for the function $\rho$. An important
about the matrix $\hat{A}$ is that it is not constant on the
world-sheet, quite in opposite to the point-particle case. On the
other hand, since an expression \bea \label{sm}
\varkappa\cosh\rho\Gamma^0-\omega_1\sinh\rho
\Gamma^2+\rho'\Gamma^3 \eea multiplied with $i$ takes values in
${\cal A}^{(2)}$, one can always find a similarity transformation
with an element $V$ from ${\rm SO}(3,1)$, which brings $\hat{A}$
to a constant matrix, e.g, to $\Gamma^0$, namely, \bea
\varkappa\cosh\rho\Gamma^0-\omega_1\sinh\rho
\Gamma^2+\rho'\Gamma^3=\omega_2 V\Gamma^0V^{-1}\, ,
\label{funny}\eea where we have taken into account that on
solutions of the Virasoro constraint (\ref{Virasoro}), the
eigenvalues of matrix (\ref{sm}) are $\pm \omega_2$. For instance,
one can take \bea V=v\left(\begin{array}{cccc}
\omega_1\sinh\rho-i\rho' & 0 & 0 & \omega_2-\kappa\cosh\rho\\
0 & \omega_1\sinh\rho+i\rho' & \omega_2-\kappa\cosh\rho & 0 \\
0 & \omega_2-\kappa\cosh\rho & \omega_1\sinh\rho-i\rho' & 0 \\
\omega_2-\kappa\cosh\rho & 0 & 0 & \omega_1\sinh\rho+i\rho'
\end{array}\right)\, ,
\eea where the unessential normalization constant is fixed by
requiring $\det V=1$. Indeed, one can check that on solutions of
the Virasoro constraint the relation (\ref{funny}) is satisfied.
Thus, the matrix $\hat{A}$ exhibits the following factorizable
structure \bea \hat{A}=\frac{\omega_2}{4}\mathscr{V}\mathscr{A}
\mathscr{V}^{-1}\, , \eea where we have introduced two matrices:
\bea {\mathscr A}=
\left(\begin{array}{cc} -i\Gamma^0 & 0 \\
0 & T_{34}+T_{56}
\end{array}
\right)\, , ~~~~~~~~~ \mathscr{V}=
\left(\begin{array}{cc} e^{-\frac{i}{2}\rho\Gamma^3}V & 0 \\
0 & {\mathbb I}
\end{array}
\right)\, , \eea where, in particular, matrix $\mathscr{A}$ does
not depend on the world-sheet variables. We thus see that under a
linearized $\kappa$-symmetry transformation the combination
${\mathscr V}^{-1}\chi \mathscr{V}$ undergoes a shift by an
element \bea \frac{\omega^2}{16}\Big[ {\mathscr A}^2({\mathscr
V}^{-1}\kappa {\mathscr V})+{\mathscr A}({\mathscr V}^{-1}\kappa
{\mathscr V}){\mathscr A}+({\mathscr V}^{-1}\kappa {\mathscr
V}){\mathscr A}^2-\frac{1}{8}{\rm str}(\Upsilon^2 {\mathscr
A}^2)({\mathscr V}^{-1}\kappa {\mathscr V}) \Big]\, . \eea An easy
calculation shows that the matrix above has a structure
\bea
\label{ks} \frac{\omega_2^2}{16}\, \, \left(\begin{array}{cc} 0
~&~ \varepsilon
\\
-\varepsilon^tC_4 ~&~ 0
\end{array} \right) , \eea
where $\varepsilon$ the matrix  $\varepsilon$ depends on 8
fermions only, i.e. the rank of the on-shell $\kappa$-symmetry
transformations is equal to eight, confirming thereby the
conclusions of \cite{Arutyunov:2008if}. Thus, our analysis shows
that transformation (\ref{ks}) suffices to gauge away from the
general element \bea \mathscr{V}^{-1} \chi\mathscr{V}=\left(
\begin{array}{cc}
0 & V^{-1}e^{-\frac{i}{2}\rho\Gamma_3}\theta \\
-(V^{-1}e^{-\frac{i}{2}\rho\Gamma_3}\theta)^tC_4 & 0
\end{array}
\right)\,  \eea
 precisely eight fermions.

\smallskip

Finally, we note that in section (\ref{QFA}) we made an additional
rotation of $\chi$ with the matrix $W$ given by eq.(\ref{Wmat}).
To find how the new fermionic matrix transforms under
$\kappa$-symmetry, we have to rotate the parameter $\kappa$ in the
same way $\kappa\to W\kappa W^{-1}$. In the $\CP^3$ sector this
rotation effectively leads to modifying the matrix ${\mathscr A}$
in the following way
$$
{\mathscr A}~~\to~~{\mathscr A}=
\left(\begin{array}{cc} -i\Gamma^0 & 0 \\
0 & T_{6}
\end{array}
\right)\, ,
$$
which is the consequence of $e^{\frac{\pi}{4}
T_5}(T_{34}+T_{56})e^{-\frac{\pi}{4}T_5}=T_6$. This new matrix
${\mathscr A}$ coincides with the one used in the paper
\cite{Arutyunov:2008if}, where it was concluded that the
corresponding $\kappa$-symmetry transformations allow one to make
the gauge choice
$$
\theta T_{56}=0\, ,
$$
which puts to zero the fifth and the sixth column of $\theta$.

\end{document}